\documentstyle[prl,twocolumn,aps,epsf]{revtex}
\begin{document}

\title { Multi-pion Bose-Einstein correlation effects on three-pion interferometry}
\author{Q.H. Zhang}
\address{Institut f\"ur Theoretische Physik, Universit\"at Regensburg,
D-93040 Regensburg, Germany}
\vfill
\maketitle

\begin{abstract}
Multi-pion correlations effect on 
 three-pion interferometry is studied. It is shown that 
multi-pion correlations decrease both the apparent 
radius of the source and the coherent source parameter 
derived from three-pion interferometry. The data 
of OPAL group are discussed.
\end{abstract}

PACS number(s): 11.38 Mh, 11.30 Rd,25.75 Dw


Hanbury-Brown and Twiss\cite{HBT} were the first who applied the Bose-Einstein (BE)
correlation to measure the size of distant stars. 
The method was first applied to 
particle physics by Goldhaber et al.(GGLP)\cite{GGLP} in 1959. Since then the size 
of the interaction region has been measured by numerous experiments in high 
energy collisions using different types of particles.  
Two-pion Bose-Einstein(BE) correlation is widely used in high energy 
collisions to provide the information of the space-time
structure, degree of coherence and dynamics of the region where 
the pions were produced\cite{GKW,ZL}. 
Experimentally, ultrarelativistic hadronic and nuclear collisions provide 
the environment for creating dozens, and in some cases hundreds, 
of pions \cite{OPAL,NA35,E802}.  Theoretically, It is easy 
to extend the two-pion correlation to three-pion  
correlation function. Because the bosonic nature of the 
pion causes the abundance of pions at low mometum, It is 
anticipated that multi-pion correlation should 
affect the single and $i$-pion spectra and distort the $i$-pion 
correlation function\cite{WC84,Zajc87,Pratt93,PGG90,CGZ95,ZCG95}. 
Thus, it is very interesting to analyse the 
multi-pion  Bose-Einstein correlation effects on $i$-pion interferometry.

The general definition of the $n$ pion correlation function 
$C_{n}(\vec{p}_{1},\cdot \cdot \cdot ,\vec{p}_{n})$ is
\begin{eqnarray}
C_{n}(\vec{p}_{1},\cdot \cdot \cdot, \vec{p}_{n})
=\frac{P_{n}(\vec {p}_{1},\cdot \cdot \cdot, \vec{p}_{n})}
{\prod_{i=1}^n P_{1}(\vec{p_{i}})}   ,
\end{eqnarray}
where $P_{n}(\vec{p}_{1},\cdot \cdot \cdot, \vec{p}_{n})$ is the 
probability of observing $n$ pions with momenta $\{ \vec{p}_{i} \}$
all in the same event.   The n-pion momentum probability 
distribution $P_{n}(\vec p_1,\cdot\cdot\cdot, \vec p_n)$ 
 can be expressed as\cite{Zajc87,Pratt93,CGZ95}
\begin{equation}
P_{n}(\vec p_1,\cdot\cdot\cdot,\vec p_n)
=\sum_{\sigma} \rho_{1,\sigma(1)}\rho_{2,\sigma(2)}...\rho_{n,\sigma(n)},
\end{equation}
with
\begin{equation}
\rho_{i,j}=\rho(p_{i},p_{j})=
\int d^{4}x  g_w(x, \frac{(p_{i}+p_{j})}{2}) 
e^{i(p_{i}-p_{j})\cdot x}.
\end{equation}
$\sigma(i)$ denotes the $i$th element of a permutation of the sequence
${1,2,3,\cdot \cdot \cdot, n}$, and the sum over $\sigma$ denotes the 
sum over all $n!$ permutations of this sequence. $g_w(x,k)$ can be 
explained as the probability of finding a pion at point $x$ with momentum 
$k$ which is defined as\cite{Pratt84,CGZ94}
\begin{equation}
g_{w}(Y,k)=\int d^{4}y  j^{*}(Y+y/2)j(Y-y/2) \exp(-ik\cdot y) .
\end{equation}
Where $j(x)$ is the 
current of the pion, which can be expressed as\cite{CGZ94,ZCG95} 
\begin{equation}
j(x)=\int d^{4}x' d^{4}p  j(x',p) \nu(x')  \exp(-ip\cdot (x-x')) .
\end{equation}
Here $j(x',p)$ is the probability amplitude of finding a pion 
with momentum $p$ , emitted by the emitter at $x'$. $\nu(x')$ 
is a random phase factor which has been taken away from $j(x',p)$.  
All emitters are uncorrelated in coordinate space when assuming:
\begin{equation}
<\nu^{*}(x')\nu(x)>=\delta^{4}(x'-x)   .
\end{equation}
For simplicity we 
also assume that
\begin{equation}
<\nu^{*}(x)>=<\nu(x)>=0    ,
\end{equation}
which means that for each emitter the phases are randomly distributed in the 
range of $0$ to $2 \pi$. Then we have the following relationship
\begin{equation}
<\nu^{*}(x')\nu^{*}(x)>=<\nu(x)\nu(x')>=0  .
\end{equation}

Inserting eq.(5) into eq.(4) we have
\begin{eqnarray}
g_{w}(Y,k)&=&\int d^{4}y  \exp(-iky)
\int d^{4}x' j^{*}(x',p_{1})d^4p_{1}
\nonumber\\
&&e^{ip_{1}\cdot (Y+y/2-x')}\nu^{*}(x')
\nonumber\\
	&&\int d^{4}x''j(x'',p_{2})d^4p_{2}e^{-ip_{2}\cdot (Y-y/2-x'')}\nu(x'') ,
\end{eqnarray}
Taking phase average and using the the relationship of eq.(6), we have
\begin{eqnarray}
g_{w}(Y,k)&=&\int d^{4}y \exp(-ik\cdot y)
\int d^{4}x'' j(x'',p_{2})d^4p_{2}
\nonumber\\
&&e^{-ip_{2}\cdot (Y-y/2-x'')} ,
\nonumber\\
	&&\int d^{4}x' j^{*}(x',p_{1})d^4p_{1} 
e^{ip_{1}\cdot (Y+y/2-x')}\delta^4(x'-x'')
\nonumber\\
&=&\int d^{4}y exp(-ik\cdot y)\int d^4x' 
\nonumber\\
&&j^*(x',p_1)j(x',p_2)d^4p_1d^4p_2
\nonumber\\
&&e^{ip_1\cdot (Y+y/2-x')}e^{-ip_2\cdot (Y-y/2-x')}\\
&=&\int d^{4}y exp(-ik\cdot y)\int d^4x'j^*(x',k'+q/2)
\nonumber\\
&&j(x',k'-q/2)d^4k'd^4q
e^{iq\cdot (Y-x')}e^{-ik'\cdot y}
\nonumber\\
&=&\int d^4x' d^4q j^*(x',k+q/2)j(x',k-q/2)e^{iq\cdot (Y-x')}
\nonumber
\end{eqnarray}
Here $k'=(p_1+p_2)/2, q=p_1-p_2$ .
From eq.(1), the two-pion and three-pion correlation function 
 can be expressed as
\begin{eqnarray}
&&C_{2}(\vec{p}_{1},\vec{p}_{2})=
1+
\nonumber\\
&&\frac{\int d^{4}x d^{4}x' g_{w}(x,k_{12})g_{w}(x',k_{12})
\exp(iq_{12}\cdot (x-x'))}{\int d^{4}x d^{4}x'g_{w}(x,p_{1})g_{w}(x',p_{2})} 
\end{eqnarray}

\begin{eqnarray}
&&C_{3}(\vec{p}_{1},\vec{p}_{2},\vec{p}_3)=
1
\nonumber\\
&&+\frac{\int d^{4}x d^{4}x' g_{w}(x,k_{12})g_{w}(x',k_{12})
\exp(iq_{12}\cdot (x-x'))}{\int d^{4}x d^{4}x'g_{w}(x,p_{1})g_{w}(x',p_{2})} 
\nonumber\\
&&+\frac{\int d^{4}x d^{4}x' g_{w}(x,k_{23})g_{w}(x',k_{23})
\exp(iq_{23}\cdot (x-x'))}{\int d^{4}x d^{4}x'g_{w}(x,p_{2})g_{w}(x',p_{3})} 
\nonumber\\
&&+\frac{\int d^{4}x d^{4}x' g_{w}(x,k_{31})g_{w}(x',k_{31})
\exp(iq_{31}\cdot (x-x'))}{\int d^{4}x d^{4}x'g_{w}(x,p_{3})g_{w}(x',p_{1})} 
\nonumber\\
&&+\frac{\int d^{4}x d^{4}y  g_{w}(x,k_{12})
\exp(-iq_{12}\cdot x)
g_{w}(y,k_{23})
\exp(-iq_{23}\cdot y)}
{\int d^{4}x d^{4}y d^4 zg_{w}(x,p_{1})g_{w}(y,p_{2})g_w(z,p_3)} 
\nonumber\\
&&\cdot \int d^4 z g_w(z,k_{31})
\exp(-iq_{31}\cdot z) +
\\
&&\frac{\int d^{4}x d^{4}y g_w(x,k_{12})
\exp(iq_{12}\cdot x)
g_{w}(y,k_{23})
\exp(iq_{23}\cdot y)}
{\int d^{4}x d^{4}y d^4 zg_{w}(x,p_{1})g_{w}(y,p_{2})g_w(z,p_3)} 
\nonumber\\
&&
\cdot \int d^4 z g_w(z,k_{31})
\exp(iq_{31}\cdot z)
 .\nonumber
\end{eqnarray}
Here $k_{ij}=(p_i+p_j)/2$ and $q_{ij}=p_i-p_j$ are mean and relative 
momentum of $p_i$ and $p_j$.  

For $n \pi$ events, the $i$-pion correlation function can be defined as
\begin{eqnarray}
C_{i}^{n}(p_1,\cdot \cdot \cdot,p_{i})&=&
\frac{P_{i}^{n}(\vec p_{1},\cdot \cdot \cdot \vec, p_{i})}{
\prod_{j=1}^{i}P_{1}^{n}(\vec p_{j})} ,
\end{eqnarray}
where $P_{i}^{n}(\vec p_{1},\cdot \cdot \cdot, \vec p_{i})$ is 
the modified $i$-pion inclusive
distribution in $n$ pion events which can be expressed as
\begin{equation}
P_{i}^{n}(\vec p_{1},\cdot \cdot \cdot, \vec p_i)=\frac{\int \prod_{j=i+1}^{n} d\vec p_{j}
P_{n}(\vec p_{1},\cdot \cdot \cdot,\vec p_{n})}{\int \prod_{j=1}^{n} d\vec p_{j}
P_{n}(\vec p_{1},\cdot \cdot \cdot,\vec p_{n})}.
\end{equation}

As $n$ increases, the calculation of the intergration given above becomes 
more and more complex.   Now we define the function $G_{i}(p,q)$ as\cite{Pratt93,CGZ95}
\begin{eqnarray}
G_{i}(p,q)&=& \int \rho(p,p_{1}) d\vec p_{1} \rho(p_{1},p_{2})
d \vec p_{2} \cdot \cdot \cdot 
\nonumber\\
&&\rho(p_{i-2},p_{i-1})d \vec p_{i-1}
\rho(p_{i-1},q).
\end{eqnarray}

From the expression of $P_{n}(\vec p_1,\cdot\cdot\cdot,\vec p_n)$ (Eq.(2)),
 the two-pion and three-pion inclusive distribution can be expressed as
\begin{eqnarray}
P_{2}^{n}(\vec p_1,\vec p_2)&=&\frac{1}{n(n-1)}\frac{1}{\omega(n)}
\sum_{i=2}^{n} [\sum_{m=1}^{i-1}
\nonumber\\
&&G_{m}(p_1,p_1)\cdot G_{i-m}(p_2,p_2)
\nonumber\\
&& +G_{m}(p_1,p_2)\cdot G_{i-m}(p_2,p_1) ]\omega(n-i)
\end{eqnarray}
\begin{eqnarray}
P_{3}^{n}(\vec p_1,\vec p_2,\vec p_3)&=&\frac{1}{n(n-1)(n-2)}\frac{1}{\omega(n)}
\sum_{i=3}^{n} [\sum_{m=1}^{i-2}\sum_{k=1}^{i-m-1}
\nonumber\\
&&G_{m}(p_1,p_1)\cdot G_{k}(p_2,p_2)
\cdot G_{i-m-k}
(p_3,p_3)
\nonumber\\
&& +G_{m}(p_1,p_2)\cdot G_{k}(p_2,p_1) \cdot G_{i-m-k}(p_3,p_3)
\nonumber\\
&&
+G_{m}(p_2,p_3)\cdot G_{k}(p_3,p_2) \cdot G_{i-m-k}(p_1,p_1)
\nonumber\\
&& +G_{m}(p_3,p_1)\cdot G_{k}(p_1,p_3) \cdot G_{i-m-k}(p_2,p_2)
\nonumber\\
&&
+G_{m}(p_1,p_2)\cdot G_{k}(p_2,p_3)
\\
&&\cdot G_{i-m-k}(p_3,p_1) 
 +G_{m}(p_1,p_3)\cdot G_{k}(p_3,p_2) 
\nonumber\\
&&\cdot G_{i-m-k}(p_2,p_1)
] \omega(n-i)
\end{eqnarray}
with
\begin{eqnarray}
\omega(n)=\frac{1}{n!}\int \prod_{k=1}^{n} d\vec p_{k} 
~P_{n}(p_1,\cdot\cdot\cdot,p_n)~~~.
\nonumber
\end{eqnarray}
The single-pion distribution is
\begin{eqnarray}
P_{1}^{n}(\vec p)=
\frac{1}{n}\frac{1}{\omega(n)}\sum_{i=1}^{n}G_{i}(p,p)\cdot \omega(n-i).
\end{eqnarray}

From the expression of eq.(19), we have
\begin{eqnarray}
\omega(n)=\frac{1}{n}\sum_{i=1}^{n} C(i)\omega(n-i)
\end{eqnarray}
with
\begin{equation}
C(i)=\int d\vec p ~  G_{i}(p,p).
\end{equation}

From the above method the two-pion and three-pion correlation 
can be calculated for $n$ pion events.   In the following, we 
will give an example to investigate 
the multi-pion correlation effects on two-pion and three-pion interferometry.  
We assume that the chaotic emitter amplitude distribution is
\begin{equation}
j(x,k)=\exp(\frac{-x_{1}^{2}-x_{2}^{2}-x_{3}^{2}}{2R_{0}^{2}})
\delta(x_{0})
exp(-\frac{k_{1}^{2}+k_{2}^{2}+k_{3}^{2}}{2\Delta_0^{2}})~~~.
\end{equation}
Where $R_0$ and $  \Delta_0 $ are parameters which 
represents the radius of the chaotic source and the 
momentum range of pions respectively. $(x_0,x_1,x_2,x_3)$ and $(k_0,k_1,k_2,k_3)$ 
are pion's coordinate and momentum respectively. 
Bringing  eq.(22) into 
eq.(10) , eq.(11) and eq.(12), we can easily get the function $g_w(x,k)$ 
\begin{eqnarray}
g_w(x,k)&=& (\frac{1}{\pi R_G^2})^\frac{3}{2}
exp(-\frac{\vec{x}^2}{R_G^2})\delta(x_0)
\nonumber\\
&& (\frac{1}{\pi \Delta_0^2})^\frac{3}{2}
\exp\{-\frac{\vec{k}^{2}}{\Delta_0^2}\},
\end{eqnarray}
two-pion interferometry
\begin{eqnarray}
C_{2}(\vec p_1,\vec p_2)&=&1+\exp
\{ -\frac{\vec{q}_{12}^{2}}{2}R_0^2 \}
\end{eqnarray}
and three-pion interferometry formula
\begin{eqnarray}
C_{3}(\vec p_1,\vec p_2, \vec p_3)&=&1+\exp
\{ -\frac{\vec{q}_{12}^{2}}{2}R_0^2 \}+
\exp \{ -\frac{\vec{q}_{23}^{2}}{2}R_0^2 \}
\nonumber\\
&&+
\exp \{ -\frac{\vec{q}_{31}^{2}}{2}R_0^2 \}
\nonumber\\
&&+
2\exp \{ -\frac{\vec{Q}^{2}}{4}R_0^2 \}
\end{eqnarray}
with 
\begin{eqnarray}
R_G^2=R_0^2+\frac{1}{\Delta_0^2},~~
Q^2=\vec{q}_{12}^2+\vec q_{23}^2 +\vec{q}_{31}^2.  
\nonumber
\end{eqnarray}

Now we consider the multi-pion correlation effects 
on two-pion and three-pion correlation function. 
$g_w(x,\frac{p+q}{2})$ can be expressed as
\begin{equation}
g_w(x,\frac{p+q}{2})=
\frac{1}{(\pi R_{G}^{2})^{3/2}}e^{-\frac{r^{2}}{R_{G}^{2}}}  
\frac{1}{(\pi \Delta_{0}^{2})^{3/2}}e^{-\frac{(\vec{p}+\vec{q})^2}{4\Delta_{0}^{2}}}
\delta(t) ,
\end{equation}
then we have(eq.(3))
\begin{eqnarray}
\rho(p,q)&=&\int g_w(x, \frac{p+q}{2})e^{i(p-q)x}dx
\nonumber\\
& =&\frac{1}{(2\pi \Delta_{0}^{2})^{3/2}}
e^{-\frac{(p-q)^{2}R_{G}^{2}}{4}}e^{-\frac{(\vec{p}+\vec{q})^2}{4\Delta_{0}^{2}}}.
\end{eqnarray}

Define
\begin{equation}
G_{n}(p,q)=\int \rho(p,p_{1}) 
\prod_{i=1,n-2} d\vec p_{i} 
\rho(p_{i},p_{i+1})
d\vec p_{n-1} 
\rho(p_{n-1},q)
\end{equation}
Using eq.(27),  we can easily get 
\begin{equation}
G_{n}(p,q)= \alpha_{n} e^{-a_{n}(p^{2}+q^{2})+g_{n} \vec p \cdot \vec q}
\end{equation}
where
\begin{eqnarray}
a_{n+1}&=&\frac{R_{G}^{2}}{4}+\frac{1}
{4\Delta_{0}^{2}}-\frac{(R_{G}^2-1/\Delta_0^2)^2}{16b_{n}},~
\nonumber\\
b_{n}&=&a_{n}+\frac{1}{4\Delta_{0}^{2}}+\frac{R_{G}^{2}}{4},~
\nonumber\\
g_{n+1}&=&\frac{g_{n}}{4b_{n}}\cdot(R_G^2-\frac{1}{\Delta_0^2})
\nonumber
\end{eqnarray}
and
\begin{equation}
\alpha_{n+1}=\alpha_{n}(\frac{1}{\Delta_{0}^{2}})^{3/2}(\frac{1}{b_{n}})^{3/2}
\end{equation}
with 
\begin{equation}
a_{1}=\frac{R_{G}^{2}}{4}+\frac{1}{4\Delta_{0}^{2}}, ~ 
g_{1}=\frac{R_{G}^{2}}{2}-\frac{1}{2\Delta_0^2} , ~ 
\alpha_{1}=\frac{1}{(\pi \Delta_{0}^{2})^{3/2}}.
\end{equation}

From the above formula, we can calculate the two-pion and three-pion 
correlation function in $n$ pion events.  
In fig.1, the multi-pion correlation effects on two-pion interferometry is
shown. It is clear that multi-pion correlation make the two-pion correlation 
function become broader and the intercept become lower\cite{AFS83,AFS87,UA189}. 
That is multi-pion correlation makes the apparent radius and 
coherent parameter derived from 
two-pion interferometry become smaller. 

\vskip -4.0cm
\begin{figure}[h]\epsfxsize=10cm
\centerline{\epsfbox{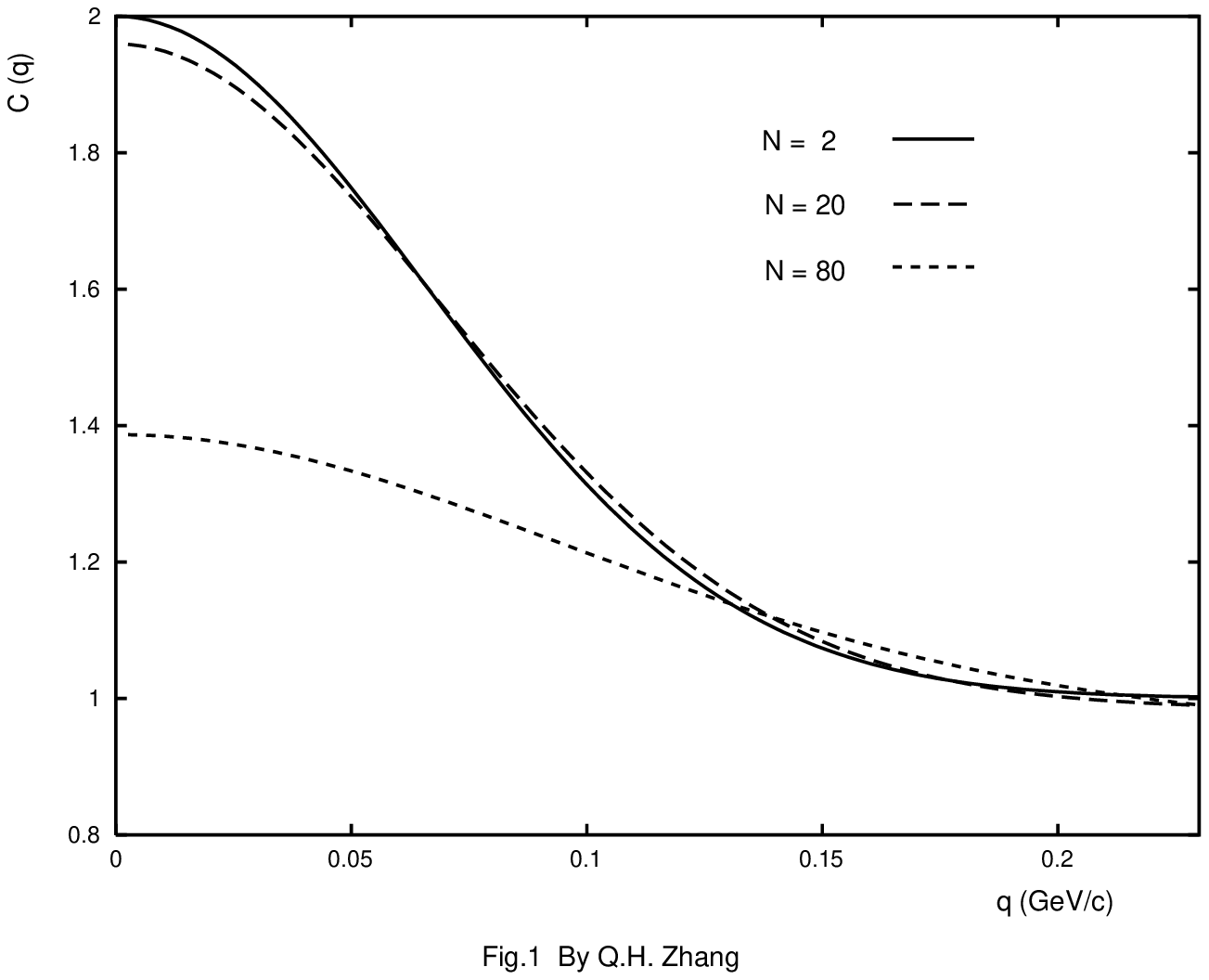}}
\vskip -2.5cm
\caption{\it 
Multi-pion correlation effects on two-pion interferometry. 
The solid line, dashed line and dotted line  
correspond to multiplicity $N=2$,$20$ and $80$ respectively. 
The input value 
of $R_0$ and $\Delta$ is $3 fm$ and $0.36 GeV$ respectively.
}
\end{figure}

For the three-pion interferometry, we choose the variable 
$Q^2=\vec{q}_{12}^2+\vec{q}_{23}^2+\vec{q}_{31}^2$ and  
integrate the other eight variables, then we 
have the three-pion correlation function $C_3^n(Q)$  
\begin{eqnarray}
&&C_3^{n}(Q)=
\nonumber\\
&&\frac{\int P_3^{n}(\vec{p}_1,\vec{p}_2,\vec{p}_3)\delta(
Q^2-\vec{q}_{12}^2-\vec{q}_{23}^2-\vec{q}_{31}^2)d\vec{p}_1 d\vec{p}_2
d\vec{p}_3}{\int P_1^{n}(\vec{p}_1)P_1^n(\vec{p}_2)P_1^n(\vec{p_3})
\delta(Q^2-\vec{q}_{12}^2-\vec{q}_{23}^2-\vec{q}_{31}^2)
d\vec{p}_1 d\vec{p}_2 d\vec{p}_3}
\end{eqnarray}
The effects of multi-pion correlation on the three-pion correlation 
function are shown in fig.2.  It can be seen clearly that as 
the multiplicity of the event increases,  
the three-pion correlation function has a lower chaoticity, though
the actual source is totally chaotic.  The three-pion 
correlation function also become broader for larger multiplicity. 
That means the apparent 
radius derived from three-pion interferometry 
becomes smaller. 
As the multiplicity increases, 
the multi-pion correlation effects on 
three-pion correlation effects becomes larger.

\vskip -4.0 cm
\begin{figure}[h]\epsfxsize=10cm
\centerline{\epsfbox{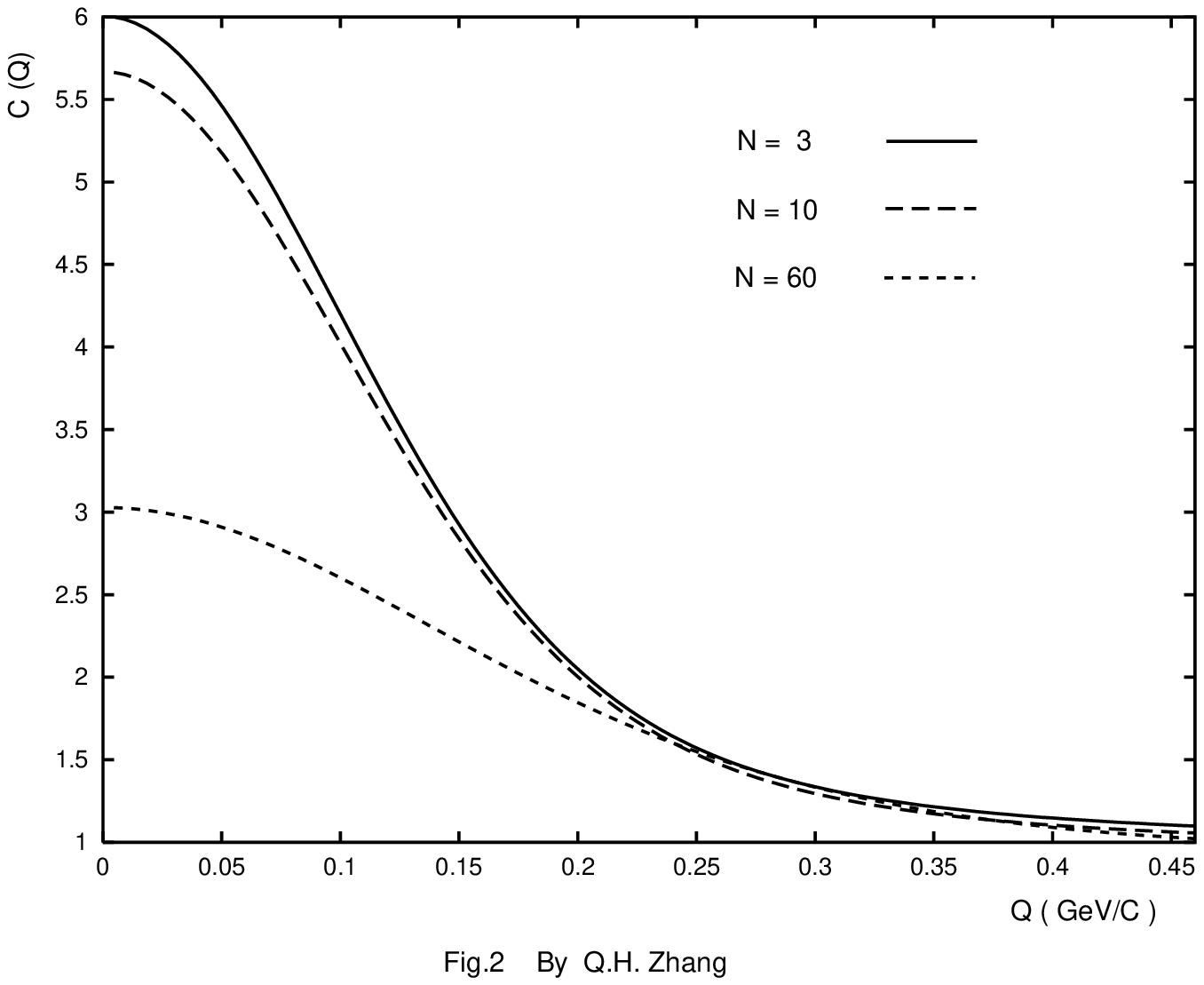}}
\vskip -2.5cm
\caption{\it 
Multi-pion correlation effects on three-pion interferometry.  
The solid line, dashed line and dotted line  
correspond to multiplicity $N=3$,$10$ and $60$ respectively. 
The input value 
of $R_0$ and $\Delta$ is $3 fm$ and $0.36 GeV$ respectively.
}
\end{figure}

Recentally, OPAL group\cite{OPAL96} have studied the source radius $R$ and 
chaoticity degree $\lambda$ as a function of charged  multiplicity. It is 
found that  
$\lambda$ decreases with 
pion multiplicity while $R$ increases with pion multiplicity. Those effects ,
though small, are statistically significant. Our calculations are not completely  
consistent with OPAL results due to the fact that: In our analysis,
we assumed that the true source radius and coherent parameter are same 
for different multiplicity which may be not true in experiment
\cite{OPAL96}. One of the 
possible explanation about the fact that $R$ increases with pion 
multiplicity (which is not consistent with our calculation) 
was given by OPAL group: At larger pion 
multiplicity, the events is dominated by three jet events which has 
larger radius than two jet events which dominate lower multiplicity events.
The larger the pion multiplicity, the larger 
the Bose-Einstein correlation effects on the radius. So It is anticipated that 
the true radius vs. pion multiplicity will become more 
steeper if we can exclude the multi-pion Bose-Einstein correlation effects. 
Due to the fact that 
a increase of the fraction of resonance at high pion multiplicity
\cite{ALEPH91}, we have another 
possible explanation: 
With resonance, the source is consisted of two parts,
one corresponds to the source consist of the direct pions 
and the pions emitted from resonance with life-time less than 
one fm, the another source is consist of the pions which decay from 
long lived resonance. The combination of a primary pion 
with a decay pion from a long lived resonance can cause the reduction of 
coherent degree $\lambda$ and increase the pion source radius
\cite{Grassberger77}. So the observed results of OPAL group may be due to 
the increasing of resonance at high multiplicities.

Conclusions: In this paper, the multi-pion Bose-Einstein correlation 
effects on two-pion and three-pion interferometry are discussed.  
It is shown that 
multi-pion Bose-Einstein correlation make the apparent
radius and coherent parameter of source which derived from 
lower order pion interferometry become smaller.  For larger pion multiplicity,
the multi-pion correlation effects on the lower order pion 
interferometry becomes larger. Recent data of OPAL group are discussed. 
It is argued that the increase of radius $R$ with pion 
multiplicity may be due to the 
increasing fraction of resonance at high multiplicities. 

\begin{center}
{\bf Acknowledgement}
\end{center}
The author would like to express his gratitude to Dr. Yang Pang and U. Heinz
 for helpful discussions.  This work was partly supported by 
the Alexander von Humboldt foundation in Germany.

\begin{center}
{\bf Figure Captions}
\end{center}


\begin{thebibliography}{29}
\bibitem{HBT}
R. Hanbury-Brown and R. Q. Twiss, Nature (London) 178 (1956) 1046.
\bibitem{GGLP}
G. Goldhaber, S. Goldhaber, W. Lee and A. Paris,  Phys. Rev. 120 (1960) 300.
\bibitem{GKW}
M. Gyulassy, S. K. Kauffmann and Lance Wilson, Phys. Rev. C 20 (1979) 2267.
\bibitem{ZL}
W. A. Zajc, in Hadronic Multiparticle Production, ed. P. Carruthers
(World Scientific, Singapore, 1987)p. 125.
\bibitem{OPAL}
OPAL Collab. P.D. Acton et al., Phys. Lett. B 320 (1994) 417
	;Phys. Lett. B267 (1991) 143.
\bibitem{NA35}
NA35 Collab..  T. J. Humanic,  Z. Phys. C38 (1988) 79.
\bibitem{E802}
E802 Collab.., T. Abbott et al., Nucl. Phys. A544 (1992) 237.
\bibitem{WC84}
W. Willis and C. Chasman, Nucl. Phys. A418 (1984) 425c.
\bibitem{Zajc87}
W. A. Zajc, Phys. Rev. D35 (1987) 3396.
\bibitem{Pratt93}
S. Pratt, Phys. Lett. B301 (1993) 159.
\bibitem{PGG90}
S. Padula, M. Gyulassy, and S. Gavin, Nucl. Phys. B329 (1990)357.
\bibitem{CGZ95}
W.Q. Chao, C.S. Gao and Q. H. Zhang, J. Phys. G21 (1995) 847.
\bibitem{ZCG95}
Q.H. Zhang,W.Q. Chao and C.S. Gao, Phys. Rev. C52 (1995) 2064.
\bibitem{Pratt84}
S. Pratt, T. Cs\"org\"o, J. Zimanyi, Phys. Rev C42 (1990)2646.
\bibitem{CGZ94}
W. Q. Chao, C. S. Gao, and Q. H. Zhang; Phys. Rev. C49 (1994) 3224.
\bibitem{AFS83}
AFS Collab., T. \AA kesson et al, Phys. Lett. B129 (1983) 269.
\bibitem{AFS87}
AFS Collab., T. \AA kesson et al, Phys. Lett. B187 (1987) 420.
\bibitem{UA189}
UA1 Collab., C. Albajar et al., Phys. Lett. B226 (1989) 410.
\bibitem{OPAL96}
OPAL Collab., G. Alexander et al., CERN-PPE/96-90, submitted to Z. Phys. C.
\bibitem{ALEPH91}
ALEPH Collaboration; D. Decamp et al.; Phys. Lett. B 273(1991) 181. 
\bibitem{Grassberger77}
P. Grassberger, Nucl. Phys. B120(1977) 231.

G. Bowler, Particle Word 2 (1991) 1.
\end{thebibliography}

\begin{enumerate}
\bibitem 1
Multi-pion correlation effects on two-pion interferometry. 
The solid line, dashed line and dotted line  
correspond to multiplicity $N=2$,$20$ and $80$ respectively. 
The input value 
of $R_0$ and $\Delta$ is $3 fm$ and $0.36 GeV$ respectively.
\bibitem 2
Multi-pion correlation effects on three-pion interferometry.  
The solid line, dashed line and dotted line  
correspond to multiplicity $N=3$,$10$ and $60$ respectively. 
The input value 
of $R_0$ and $\Delta$ is $3 fm$ and $0.36 GeV$ respectively.
\end{enumerate}
\end {document}